\documentclass[reprint,secnumarabic,amssymb,nobibnotes,aps,prl,groupedaddress,twocolumn]{revtex4-1}

\usepackage{graphicx}
\usepackage{dcolumn}
\usepackage{bm}
\usepackage{siunitx}
\usepackage[dvipsnames]{xcolor}
\usepackage{lipsum}
\usepackage{verbatim}
\usepackage{soul}
\usepackage{amsmath}
\usepackage{comment}
\usepackage{comment}
\usepackage[normalem]{ulem}

\makeatletter
\newif\ifpreprint
\@ifclasswith{revtex4-1}{preprint}{\preprinttrue}{\preprintfalse}
\makeatother

\begin{document}

	\title{The electron cyclotron maser instability in laser-ionized plasmas}
	
	\author{Thales Silva}%
	\email{thales.silva@tecnico.ulisboa.pt}
	\affiliation{GoLP/Instituto de Plasmas e Fus\~ao Nuclear, Instituto Superior T\'ecnico, Universidade de Lisboa, 1049-001 Lisbon, Portugal}
	\author{Pablo J. Bilbao}%
	\affiliation{GoLP/Instituto de Plasmas e Fus\~ao Nuclear, Instituto Superior T\'ecnico, Universidade de Lisboa, 1049-001 Lisbon, Portugal}
	\author{Luis O. Silva}%
	\email{luis.silva@tecnico.ulisboa.pt}
	\affiliation{GoLP/Instituto de Plasmas e Fus\~ao Nuclear, Instituto Superior T\'ecnico, Universidade de Lisboa, 1049-001 Lisbon, Portugal}
	\date{\today}
	
	\begin{abstract}
		We show that circularly polarized lasers create plasmas with long-lasting ring-shaped weakly relativistic momentum distributions which, in the presence of an ambient magnetic field, are prone to the electron cyclotron maser instability. Theoretical results and particle-in-cell simulations show that current laser technology can effectively induce field ionized tailored distribution functions and probe the electron cyclotron maser in controlled conditions, providing direct experimental evidence to coherent radiation processes driven by ring-shaped or Landau inverted momentum distributions of relevance in extreme astrophysical conditions.
	\end{abstract}
	
	\maketitle
	
	The electron cyclotron maser (ECM) is one of the most prominent mechanisms for coherent radiation generation in space and astrophysical plasmas \cite{1982Melrose, 1985Winglee, 2004Chu, 2006Treumann, 2017Melrose}. These plasmas are often non-Maxwellian and magnetized; under these conditions, the ECM has been invoked as the source of radiation for a wide range of phenomena, from auroral kilometric radiation \cite{1979Wu,1984Pritchett,1998Delory,2000Bingham} to extreme astrophysical objects such as magnetars \cite{2019Metzger, 2022Khangulyan}, blazars \cite{2005Begelman}, and astrophysical shocks \cite{2003Bingham, 2019Plotnikov, 2019Speirs, 2021Sironi}. The nature of these systems limits our ability to obtain precise, \textit{in-situ} measurements or to understand the key elements that explain the radiation signatures. In these extreme scenarios, there are many open questions, including the multi-dimensional behavior of the plasma, the competition between different plasma modes, and the long temporal evolution (and saturation) of the ECM. Investigating the ECM as a source of coherent radiation from extreme astrophysical objects is also especially timely, given recent findings that plasmas unstable to the ECM instability are a general feature of Vlasov-Maxwell dynamics when considering radiation reaction in strong magnetic fields \cite{2023Zhdankin, InPrepPablo, 2024Bilbao, 2024Ochs, 2024Bilbao_2}.
	
	Mimicking astrophysical relevant phenomena or microphysics under controlled conditions is one of the most exciting frontiers in laboratory astrophysics \cite{2006Remington, 2019Lebedev}. For phenomena that emerge on plasma kinetic scales, such as the ECM, shaping the plasma momentum distribution function (distribution function) is essential \cite{2021Silva}. Numerous configurations have been explored, using interpenetrating plasma flows, laser-plasma interactions, or beam-plasma interactions have been proposed to generate shaped unstable distribution functions \cite{2011Huntington, 2012Allen, 2012Fiuza, 2012Shukla, 2013Fox, 2015Huntington, 2017Gode, 2017Nerush, 2018Shukla, 2019Zhang, 2020Silva, 2020Raj, 2020Shukla, 2020Shukla_2, 2022Shukla, 2022Boella, 2024Huang}. These studies have mostly focused on unmagnetized plasmas, and only recently have magnetized scenarios started to be considered \cite{2023Perez}. Ambient magnetic fields, prevalent in many astrophysical scenarios, introduce different instabilities \cite{1983Galeev} and can suppress those found in unmagnetized plasmas \cite{1975Molvig}. Among all the different techniques to generate shaped distribution functions, the most easily generalized to magnetized scenarios relies on using intense laser pulses to tunnel-ionize neutral gases \cite{2019Zhang, 2024Huang}.
	
	Laser pulses can imprint their polarization on the distribution function during the optical field ionization process \cite{1992Leemans, 2019Zhang, 2020Zhang, 2022Zhang}. This arises from the conservation of canonical momentum $\mathbf{P}$ for the ionized electrons, given by $\mathbf{P} = \mathbf{p} - e \mathbf{A}/c$, where $\mathbf{A}$ is the laser vector potential,  $e$ is the elementary charge and $c$ is the speed of light in vacuum). Since the linear momentum $\mathbf{p}$ at the ionization instant is negligible (ionization electrons are born at rest), the canonical momentum after the passage of the laser is determined by the vector potential at the ionization instant $\mathbf{A}_{ion}$ \cite{2022Xu}. After the laser leaves the interaction region, the particle momentum is $\mathbf{p}^* = -e\mathbf{A}_{ion}/c$. For circularly polarized lasers, this leads to ring-like distribution functions \cite{2001Soldatov, 2001Kuzelev}. If immersed in a background guiding magnetic field, these distributions are unstable and a known source of ECM radiation. In the presence of an ambient magnetic field, the conservation of canonical momentum must be modified, and, after the laser passes, the particle momentum is $p_0 = |\mathbf{p}^*/ (1 \pm\,\Omega_c/\gamma_0\omega_0)|$ (see Appendix), where $p_0$ is the transverse momentum magnitude, and, assuming the laser propagates parallel to the magnetic field, the + and -- signs describe left and right-handed laser polarization, respectively, $\omega_0$ is the laser frequency, $\Omega_c = e B_0 /m_e c$ is the electron cyclotron frequency, $B_0$ is the intensity of the ambient magnetic field, $m_e$ the electron mass, and $\gamma_0 = (1 + p_0^2/m_e^2c^2)^{1/2}$. As long as $\Omega_c / \gamma_0\omega_0 \ll 1$, or equivalently $B_0\, (\SI{}{T})\times\lambda\, (\SI{}{\micro\meter}) \ll 10^4$, where $\lambda= 2 \pi c/\omega_0$ is the laser wavelength, the particle momentum after ionization remains approximately $p_0 \simeq |e\mathbf{A}_{ion}/c|$.
	
	In this Letter, we show that plasmas field-ionized by circularly polarized laser pulses in the presence of an external magnetic field are prone to the ECM. We uncover the competing instabilities and the parameter regimes ideal for probing radiation generated by the ECM mechanism. We present a model to predict the distribution function generated by laser ionization, determining how different gases and lasers generate distinct and controllable distribution functions. We perform multi-dimensional particle-in-cell (PIC) simulations with the OSIRIS framework \cite{2002Fonseca} to illustrate the properties of the distribution function generated in several scenarios, confirming the onset and generation of ECM radiation in a self-consistent scenario. This configuration opens a new path to studying and measuring ECM radiation in the laboratory, resorting to optical field ionized plasmas. 
	
	\begin{figure}[t]
		\centering
			\includegraphics[width=.99\linewidth]{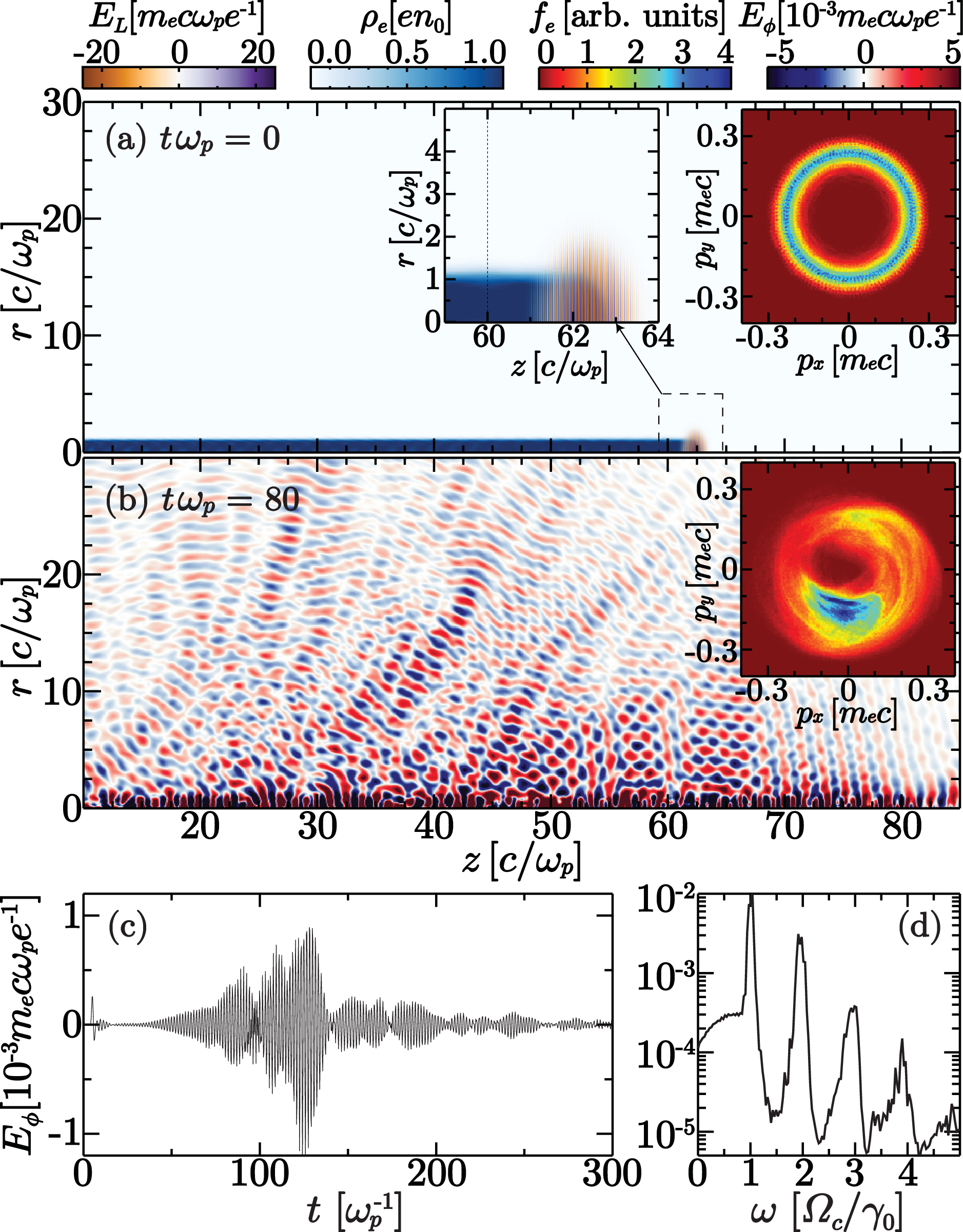}
		\caption{Simulation depicting the plasma generation with a ring distribution function and emission due to the ECM instability. (a) Circularly polarized laser ionizing a neutral helium gas generates a plasma column with radius $r= c/\omega_p = \SI{150}{\micro\meter}$, as shown in the left inset. An external magnetic field $B_0 = \SI{50}{T}$ is present in the $z$ direction. At $z = 60\,c/\omega_p$, indicated by the dotted line in the left inset, we measure the distribution function, resulting in the ring-shaped distribution function shown in the right inset, with a radius $p_0/m_ec = 0.25$. (b) At time $\tau = 80\,\omega_p^{-1} = \SI{35}{\pico\second}$, we observe modes propagating inside the plasma and modes escaping the plasma transversely. Measuring the distribution function at the same position as in panel (a) reveals azimuthal bunching, a signature of the electron cyclotron maser. (c) Temporal profile of radiation detected at $(r, z) = (30\,c/\omega_p,\, 60\,c/\omega_p)$ and (d) the frequency content of this radiation.
		}
		\label{fig:mas_sim}
	\end{figure}
	We start by summarizing our concept by performing quasi-3D \cite{2015Davidson}  self-consistent particle-in-cell simulations. Field ionization is modeled using the ADK model \cite{1986Ammosov, 2003Bruhwiler}. In Figure \ref{fig:mas_sim}(a), a circularly polarized laser, with peak normalized vector potential $\mathrm{a}_0 = e|\mathbf{A}_0|/m_ec^2 = 0.35$, $\lambda =\SI{10}{\micro\meter}$, and beam waist $w_0 = \SI{200}{\micro \meter}$, ionizes a column of neutral helium gas, producing a uniform density plasma, with peak electron density $n_e = \SI{1.5e15}{cm^{-3}}$, behind the laser. A guiding magnetic field  $\mathbf{B} = B_0\hat{e}_z$ is also present, such that $\Omega_c/\omega_p = 4$, where $\omega_p = (4\pi n_e e^2/m_e)^{1/2}$ is the electron plasma frequency.
	The plasma distribution function shows the ring structure in momentum space [right inset in \ref{fig:mas_sim}(a)], 
	and collective effects dominate the plasma dynamics after ionization; 
	Fig. \ref{fig:mas_sim}(b) reveals the active presence of electromagnetic waves propagating inside the plasma along the magnetic field direction---circularly polarized R waves \cite{1973Krall}---as well as waves born inside the plasma that escape transversely--- nearly linearly polarized X waves \cite{1973Krall}. The azimuthal bunching in momentum space, a signature of the electron cyclotron maser, is shown in the inset in Fig. \ref{fig:mas_sim}(b). The main characteristics of the X waves are illustrated in Figs. \ref{fig:mas_sim}(c) and (d): a pulse of radiation ($I = \SI{2e6}{W/cm^2}$ and duration $\tau\omega_p \approx 150 = \SI{70}{\pico\second}$), with several harmonics of the (relativistic corrected) cyclotron frequency $\Omega_c/\gamma_0  = \SI{8.5}{\tera Hz}$ are measured outside the plasma. The presence of these cyclotron harmonics is a characteristic of the X-waves and represents a measurable quantity for experimental confirmation of the ECM.
	
	Let us determine the conditions for observing ECM radiation in an optical field ionized plasma in a magnetized scenario. The onset of the ECM instability requires the plasma distribution function $f_0$ to have inverted Landau populations in some region of phase-space, i.e., $\partial f_0/\partial p_\perp >0$, where $p_\perp = (p_x^2+p_y^2)^{1/2}$ represents the momentum component perpendicular to the ambient magnetic field. Laser ionization with circularly polarized pulses generate ring-shaped distribution functions \cite{2001Kuzelev,2001Soldatov} that satisfy this requirement, with
	\begin{equation}
		f_0(p_\perp,p_z) \propto \mathrm{e}^{-(p_\perp-p_0)^2/2m_eT_\perp}\, \mathrm{e}^{-p_z^2/2m_eT_z},
		\label{eq:f0}
	\end{equation}
	where $T_\perp$ and $T_z$ denote the ring and longitudinal temperatures, respectively. Another key component for the ECM to operate is the ambient magnetic field; in its absence, the distribution in Eq. \eqref{eq:f0} becomes isotropic due to collective plasma processes \cite{2019Zhang} on a timescale $t \sim \mathcal{O}(1/\Gamma_{\text{Weibel}})$, where $\Gamma_{\text{Weibel}}$ is the maximum growth rate of the Weibel instability \cite{1959Weibel}. The ambient magnetic field induces two principal plasma modes and suppresses the Weibel mode \cite{1975Molvig}, thus allowing the ring distribution to survive over longer timescales.
	
	We focus on two magnetized modes: the slow branch of the R waves (or slow modes) and the X waves. Both modes are unstable for broad parameters if the magnetization is sufficiently large [$\Omega_c/\omega_p \gtrsim 1$]. The slow modes (with wavevector $\mathbf{k} \parallel \mathbf{B}$) are an extension of the Weibel instability into the magnetized domain, arising from the anisotropy of the ring-shaped $f_0$ [Eq. \eqref{eq:f0}], between the parallel and the perpendicular components, and tending to isotropize the distribution function \cite{1978Chu}. The unstable X waves ($\mathbf{k} \perp \mathbf{B}$) arise from intrinsically relativistic wave-particle interactions involving the particles in phase-space regions where $\partial f_0/\partial p_\perp > 0$. Electrons will azimuthally bunch in momentum space \cite{1978Chu}, producing coherent radiation. Since the X waves result from the electron cyclotron maser mechanism, the necessary condition for radiation is that the timescale for X wave generation is at least comparable to the timescale for isotropization due to the slow modes, allowing azimuthal bunching to occur effectively.
	
	\begin{figure}[t]
		\centering
		\includegraphics[width=.99\linewidth]{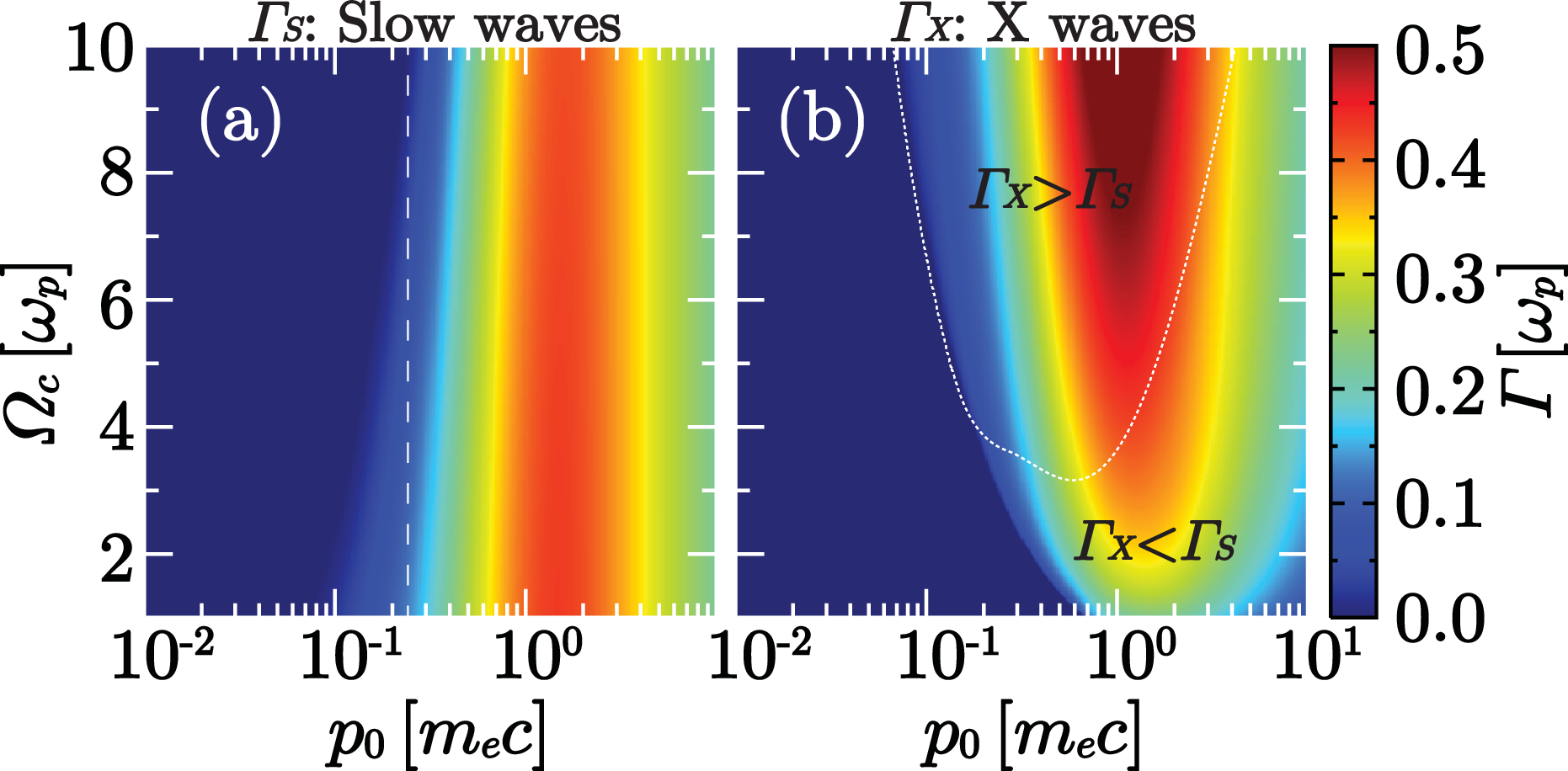}
		\caption{Theoretical maximum growth rate of the slow and X waves as a function of the parameter $p_0$ and $\Omega_c/\omega_p$, for fixed $T_z = \SI{100}{eV}$. (a) Growth rate of the slow waves increases. The vertical dashed brings attention to the suppression of the growth rate with an increasing magnetic field. (b) Growth rate of the X waves. The dotted line represents the contour where $\Gamma_X = \Gamma_S$. For significant growth of the X waves, $p_0$ and $\Omega_c/\omega_p$ should lie near this line or within the region where $\Gamma_X > \Gamma_S$.}
		\label{fig:mas_theory}
	\end{figure}
	Kinetic theory can be used to calculate the growth rates for these modes, employing the distribution function in Eq. \eqref{eq:f0}, with details provided in the Appendix. We can simplify our analysis for  $T_\perp\rightarrow0$ (the typical values of $T_\perp$ observed in simulations of laser-ionized plasmas yield theoretical results comparable to those of a cold ring).  Denoting the maximum growth rates $\Gamma_S = \Gamma_S(p_0, \Omega_c/\omega_p, T_z)$ and $\Gamma_X = \Gamma_X(p_0, \Omega_c/\omega_p)$ for the slow and the X waves, respectively, we show their dependence on the ring parameter $p_0$ that characterizes $f_0$ and the cyclotron frequency $\Omega_c$ in Figure \ref{fig:mas_theory} (for fixed $T_z=\SI{100}{eV}$, a reference value for results of our laser-ionization simulations). The growth rate of the slow modes, Fig. \ref{fig:mas_theory}(a), displays the expected behavior: it increases with $p_0$, as a larger ring-shaped distribution function enhances the anisotropy and decreases with increasing magnetic fields. Conversely, Fig. \ref{fig:mas_theory}(b) shows that the growth rate of the first harmonic of the X waves increases with both $\Omega_c$ and $p_0$. Notably, the growth rates of both modes decrease due to relativistic effects when $p_0 \gtrsim m_e c$. The absence of unstable X waves for $p_0 \lesssim 0.1\,m_ec$ demonstrates that the plasma must be weakly relativistic for the onset of the ECM. The range of parameters for X waves to dominate over R waves is determined by the condition $\Gamma_X > \Gamma_S$ (the boundary illustrated by the dotted line in Fig. \ref{fig:mas_theory})(b), yielding $\Omega_c/\omega_p \gtrsim 3$ for weakly relativistic laser-ionized plasmas with $p_0/m_ec \sim \SIrange[]{0.2}{1}{}$.
	
\begin{figure}[t]
		\centering
			\includegraphics[width=.99\linewidth]{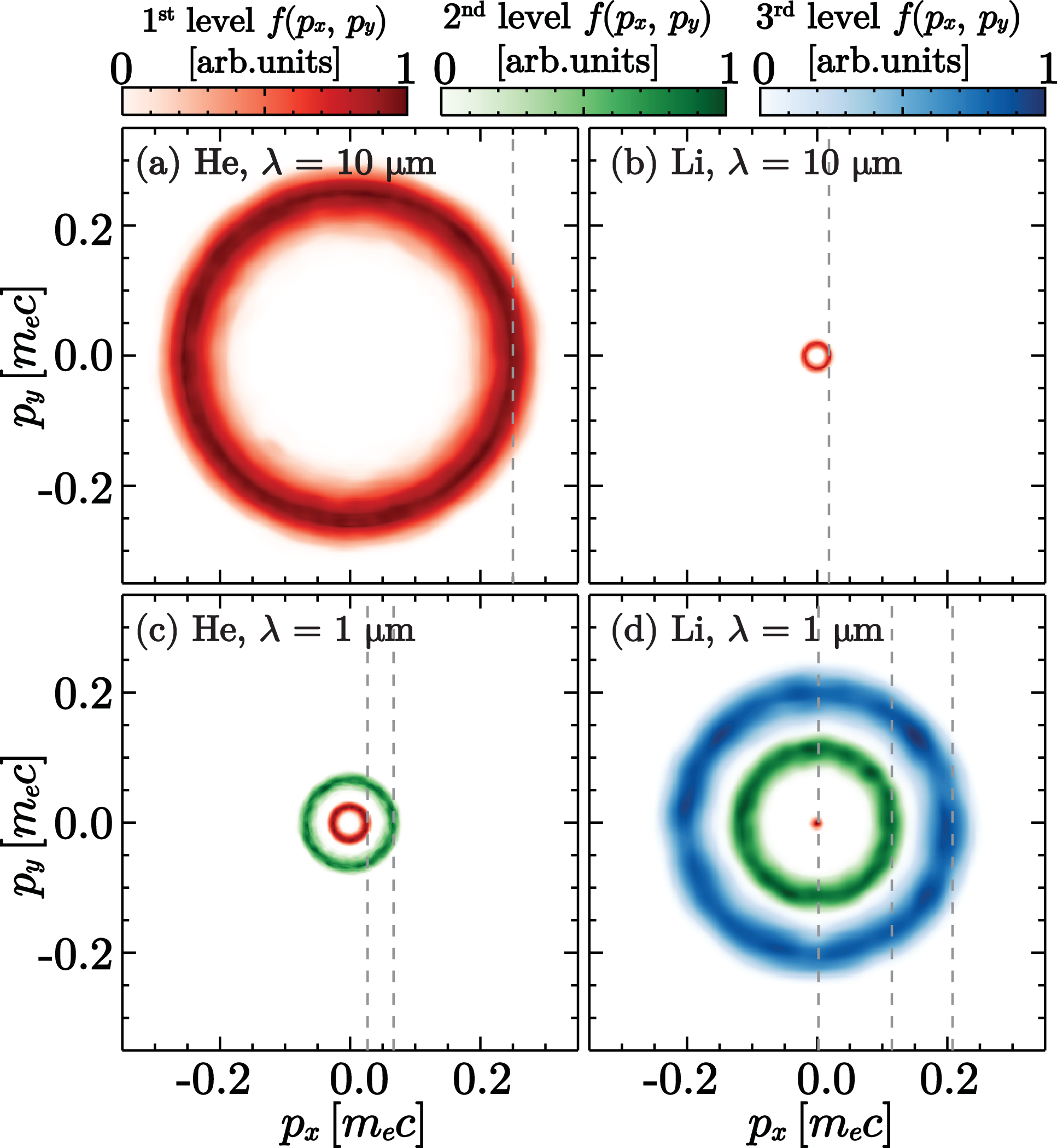}
		\caption{Ring-shaped distribution function obtained from quasi-3D particle-in-cell simulations for helium (left column) and lithium (right column) when $\lambda = \SI{10}{\micro\meter}$ (top row) and $\lambda = \SI{1}{\micro\meter}$ (bottom row). The dashed gray lines represent $p_x = p_0$ with $p_0$ predicted by Eq. \eqref{eq:p0_prediction}.}
		\label{fig:ring_gen}
	\end{figure}

	We now demonstrate that optically field-ionized plasmas can reach the desired parameter space for accessing the ECM-dominated regime. The radius $p_0$ of the ring-shaped distribution function, resulting from a circularly polarized laser, is determined by the laser vector potential at the ionization instant as $p_0/m_ec \approx \mathrm{a}_{\text{ioniz}}$, being related to the laser electric field at the time of ionization $E_{\text{ioniz}}$ via $\mathrm{a}_{\text{ioniz}} = e E_{\text{ioniz}}\lambda/2\pi m_e c$. Using the ADK model \cite{1986Ammosov, 2003Bruhwiler} to estimate the ionization rates $w(E,\xi_a, Z_a)$ depending on the local electric field $E$, the ionization energy $\xi_a$, and the atomic number $Z_a$, the fraction of ionized atoms from the $a$-th level is $n_{a}(t)/n_0 =\int_{-\infty}^{t}w(E_{\text{env}}(t^\prime),\xi_a, Z_a)dt^\prime$, where $E_{\text{env}}(t)$ is the longitudinal envelope of the laser's electric field. Full ionization of the $a$-th level of the gas (going from $n_a/n_0=0$ to $n_{a}/n_0=1$) typically occurs rapidly as $E_{\text{env}}(t)$ becomes sufficient to tunnel-ionize the neutral gas. Thus, we assume that $w(E(t),\xi_a, Z_a) \approx w(E_{\text{ioniz}},\xi_a, Z_a)=\mathrm{const.}$, during ionization, and calculate the electric field $E_{\text{ioniz}}$ required to satisfy $w(E_{\text{ioniz}}, \xi_a, Z_a)\Delta t_{\text{ioniz}} = 1$ over a time $\Delta t_{\text{ioniz}}$, a fraction of the laser duration $\tau$. Equation \eqref{eq:p0_prediction} depends weakly on $\Delta t_{\text{ioniz}}$ in a range $\Delta t_{\text{ioniz}}\sim\tau/100$ to $\tau/10$ (we choose $\tau/20$ as a figure of merit). Solving for $E_{\text{ioniz}}$ yields the distribution function ring radius $p_0/m_ec \approx \mathrm{a}_{\text{ioniz}}$
	\begin{equation}
		\frac{p_0}{m_ec} \approx - \left[ \frac{\beta\lambda(\SI{}{\micro m})}{3210.7} \right] \left[ (2 \eta - 1) W_{-1}\left(- \frac{\alpha^{\frac{1}{2 \eta - 1}} \beta}{2 \eta - 1} \right) \right]^{-1},
		\label{eq:p0_prediction}
	\end{equation} 
	where $\beta = 6.83 \,[\xi_a(eV)]^{\frac{3}{2}}$, $\eta = 3.69\,Z_a \left[\xi_a(eV)\right]^{-\frac{1}{2}}$, $\alpha =13.5\,\left[\tau(\SI{}{\femto\second})\right]^{-1}\eta\,1681^{-\eta}\,\Gamma\left(2\eta\right)  [\xi_a(eV)]^{-3 \eta+\frac{1}{2}}$, $W_{-1}$ is the Lambert function, and $\Gamma$ is the Gamma function. Equation \eqref{eq:p0_prediction} also determines the required normalized laser intensity to ionize the plasma, i.e., $\mathrm{a}_{0} \geq p_0/m_ec$. Figure \ref{fig:ring_gen} compares Eq. \eqref{eq:p0_prediction} with simulations for two gases (He and Li) and two wavelengths ($\lambda = \SI{1}{\micro\meter}$ and $\lambda = \SI{10}{\micro\meter}$), observing excellent agreement between the predicted $p_0$ and the simulations in all scenarios and for all ionization levels. Figure \ref{fig:ring_gen}(a) and (b) show that different gases may lead to very distinct regimes for the same laser parameters. For He, with higher ionization potential for the first level, the ring radius can be one order of magnitude larger than for Li. For $\lambda = \SI{10}{\micro\meter}$ and $\mathrm{a}_0 = 0.35$, the laser cannot ionize the second level of either gas. For the same $\mathrm{a}_0$, all levels of helium and lithium are ionized when $\lambda = \SI{1}{\micro\meter}$ [Figs. \ref{fig:ring_gen}(c) and (d)], but smaller rings are generated. We conclude that larger wavelengths or gases that are harder to ionize are preferable to reach regimes of $p_0/m_ec > 0.2$, where the X waves are more easily excited.
	
	\begin{figure}[t]
		\centering
			\includegraphics[width=.99\linewidth]{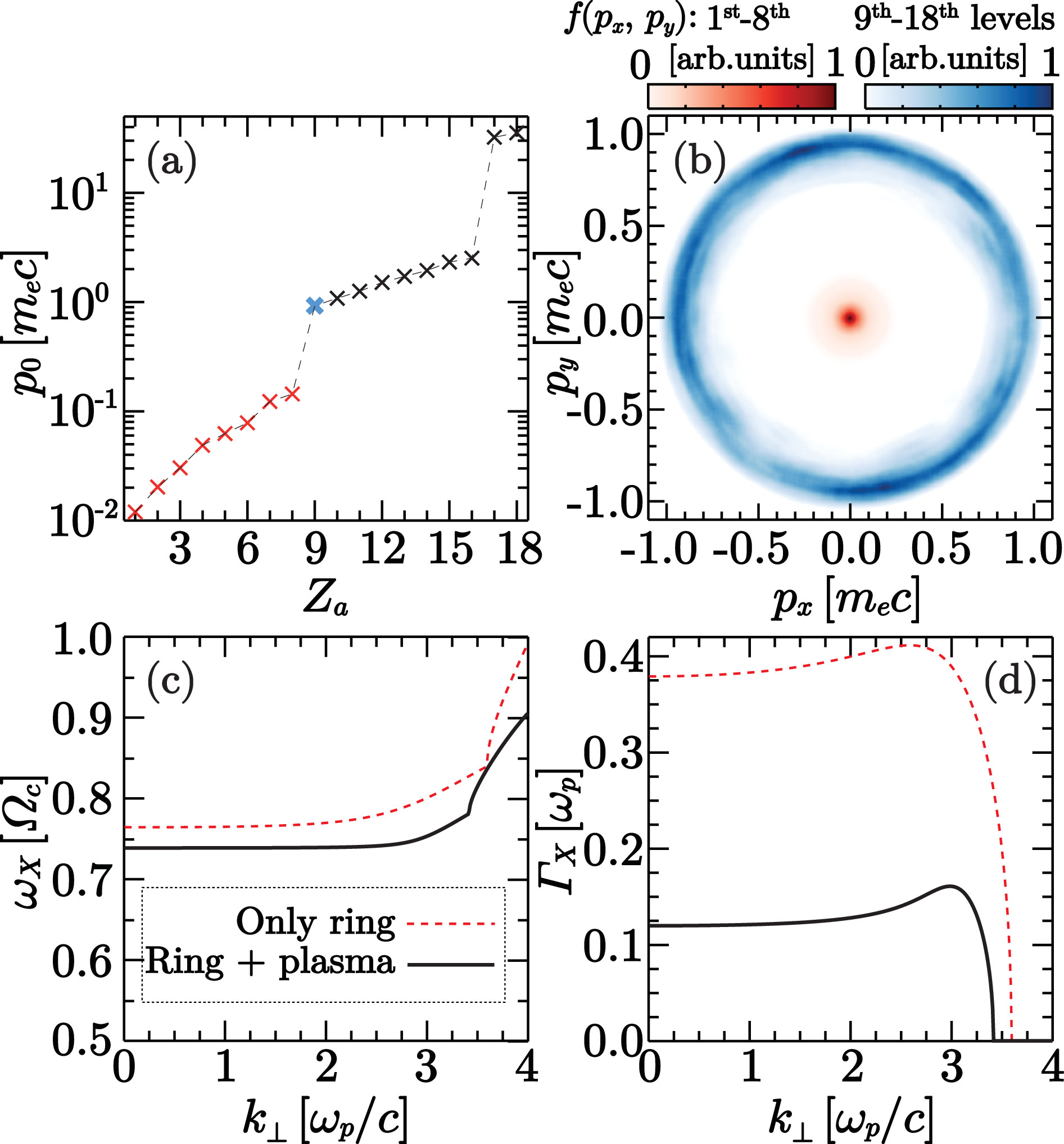}
		\caption{Demonstration of the ECM instability properties when the distribution function comprises a background population and ring for argon gas and $\lambda = \SI{1}{\micro\meter}$. (a) Prediction of the distribution function ring radius for the eighteen ionization levels of argon. (b) Simulation result for a laser with intensity $\mathrm{a}_0 = 0.95$, near the threshold for ionization up to the ninth level, showing a central population (in red) and an outer ring (blue) compatible with our prediction. Panels (c) and (d) compare solutions of the dispersion relation for varying wavevector in a case of only a ring distribution and another with $1/9$ of the population forming the ring and $8/9$ a cold plasma.}
		\label{fig:mas_800nm}
	\end{figure}
	From Fig. \ref{fig:ring_gen}(d), higher-order ionization can generate larger rings, enabling access to the relativistic regime. An appropriate combination of gases and lasers triggers the maser instability as if only for a single ring-shaped distribution function. Figure \ref{fig:ring_gen}(a) illustrates this for $\lambda = \SI{1}{\micro\meter}$ laser and $\mathrm{a}_0 = 0.95$ [the others parameters as in Fig. \ref{fig:mas_sim}] propagating in neutral Ar with an atomic density of $n=\SI{1.6e14}{\centi\meter^{-3}}$. This ensures $n_e$ for $\text{Ar}^{9+}$ remains $n_e = \SI{1.5e15}{cm^{-3}}$. Our model predicts that the first eight ionization levels will have ring radii $p_0/m_ec \leq 0.1$ [see Fig. \ref{fig:ring_gen}(a)]; the electrons from these levels form a central population which is stable. The laser ionizes up to the ninth level of Ar, Fig. \ref{fig:mas_800nm}(a), with the electron population of the last level shown in blue in Fig. \ref{fig:mas_800nm}(b). Using kinetic theory, we compare the unstable X-wave spectra for a single ring with $p_0/m_ec = 0.92$---the radius of the blue population in Fig. \ref{fig:mas_800nm}(b)---to those including both this ring and the remaining cold electrons [red in Fig. \ref{fig:mas_800nm}(b)]. Figures \ref{fig:mas_800nm}(c) and (d) show the frequency of the first harmonic of the X-waves and the growth rate as a function of the wavevector, respectively. Although the growth rate is lower by a factor of approximately $1/3$, due to the background plasma population generated by the $\mathrm{Ar}$ species up to $\mathrm{Ar}^{8+}$ the unstable wavevectors and wave frequencies are nearly identical in both cases.
	
	These two nearly independent elements allow us to generalize the scenarios unstable to the ECM for a wide range of parameters. The first element is the generation of the ring-shaped distribution function, which is controlled by the gas species and the laser---as discussed in Figs. \ref{fig:ring_gen} and \ref{fig:mas_800nm}---but is nearly independent of the gas (plasma) density. The second element is the ratio $\Omega_c/\omega_p$; as demonstrated in Fig. \ref{fig:mas_theory}, the X waves become relevant when $\Omega_c/\omega_p \gtrsim 3$. Therefore, the gas density can be adjusted for a given magnetic field to achieve the ECM regime. These two components are lightly connected, as a sufficiently wide plasma column must be created to efficiently amplify the X waves inside the plasma before they escape. By lowering $n_e$, the skin depth $c/\omega_p \propto n_e^{-1/2}$ decreases, and the laser parameters (intensity or waist) to generate wider plasma columns $ \approx c/\omega_p$.
	
	We emphasize that other regimes or phenomena can be explored with similar optically field ionized plasmas in an ambient magnetic field. For example, we conjecture that working at $\Omega_c/\omega_p < 1$, the unstable slow waves can be used to control the onset of turbulence, raising the possibility of studying the properties of these turbulent systems. In these lower magnetization conditions, the firehose and mirror instabilities can be excited with different laser polarizations. These configurations will be addressed in future publications.
	
	We have demonstrated the emission of X waves due to the electron cyclotron maser in laser-ionized plasmas through multi-dimensional particle-in-cell simulations. We have quantified how ring-shaped momentum distributions generated by circularly polarized lasers depend on the laser and gas parameters. Using kinetic theory, we have compared the growth rates of X waves with those of competing slow waves, determining the ideal regimes for observing X waves. We have shown how multi-level ionization enables access to the relativistic regime, which is critical for mimicking regimes closer to astrophysical plasmas. Recent findings suggest that ring-shaped distribution functions may occur frequently in astrophysical scenarios \cite{InPrepPablo, 2019Plotnikov, 2024Arno}, establishing the ECM as a possible mechanism for coherent radiation emitted by astrophysical objects \cite{2024Bilbao_2}. This configuration provides a timely method to explore key ECM properties and radiation signatures in the laboratory, as we have demonstrated that it is achievable using optical-field-ionized plasmas with high-intensity lasers.
	\begin{acknowledgments}
		This work was partially supported by the Portuguese Foundation for Science and Technology (FCT) through grants 2022.02230.PTDC (X-Maser), UI/BD/151559/2021, and IPFN-CEEC-INST-LA3/IST-ID. We acknowledge CINECA, the LUMI consortium, and the EuroHPC Joint Undertaking for awarding this project access to the supercomputers Leonardo (hosted at CINECA, Italy) and LUMI (hosted by CSC, Finland). We also acknowledge the National Advanced Computing Network (RNCA) for granting us access to the supercomputer MareNostrum 5 (hosted at BSC, Spain).
	\end{acknowledgments}

	\appendix
	\section{Appendix}
	\subsection{Momentum in magnetized ionization scenarios}
	The Hamiltonian for an electron interacting with a right-handed circularly polarized plane wave in the presence of a uniform magnetic field along the $\hat{e}_z$ direction is given by
	\begin{multline}
		H = \left[1 + \left[P_x - \mathrm{a}_0 f(t) \sin(t )\right]^2 \right. \\
		\left. + \left[P_y + \Omega_c x + \mathrm{a}_0 f(t) \cos(t)\right]^2 \right]^{1/2},
	\end{multline}
	where \( \mathbf{P} = P_x \hat{e}_x + P_y \hat{e}_y\) is the canonical momentum, $f(t)$ is the laser temporal profile, and we are working in normalized units (\( H \), \( \mathbf{P} \), \( t \), \( z \) are in units of \( m_e c^2 \), \( m_e c \), \( \omega_0^{-1} \), and \( c/\omega_0 \), respectively). We assume that \( \mathrm{a}_0 < 1 \), so the longitudinal motion can be neglected (\( P_z = 0 \)), and we consider the motion of an electron ionized at \( z = 0 \). Using Hamilton's equations, we solve the evolution of the linear momentum $p_x\equiv \gamma\dot x$ and $p_y\equiv \gamma\dot y$ assuming that $\gamma = H$ is nearly constant, consistently with \( \mathrm{a}_0 < 1 \) assumption, resulting in
	\begin{align}
		p_x(t) &= \mathcal{P} \cos\left(\frac{\Omega_c}{\gamma} t + \phi\right) - \frac{\mathrm{a}_0 f(t)}{1 - \Omega_c/\gamma} \sin(t), \\
		p_y(t) &= \mathcal{P} \sin\left(\frac{\Omega_c}{\gamma} t + \phi\right) + \frac{\mathrm{a}_0 f(t)}{1 - \Omega_c/\gamma} \cos(t).
	\end{align}
	At the ionization instant ($t=t_\text{ioniz}$), we assume that $p_x=p_y=0$, which allows us to determine $\mathcal P = \left|\mathrm{a}_0f(t_\text{ioniz})/(1-\Omega_c/\gamma\right)|$. Note that, in the absence of the magnetic field, the momentum after the laser passage would be $|\mathbf{p}^*/m_ec|=\mathrm{a}_0f(t_\text{ioniz})$. Thus, for large $t$ (when $f(t) \approx 0$), 
	\begin{equation}
		\left(p_x^2+p_y^2\right)^{1/2}={p_0} = \left|\frac{\mathbf{p}^*}{1-\Omega_c/\omega_0\gamma}\right|, \label{eq:radius_mag}
	\end{equation}
	where we explicitly displayed the dependence on the laser frequency. In the case of left-handed circular polarization, one can follow the same procedure and obtain a positive sign in the denominator of Eq. \eqref{eq:radius_mag}.
	\subsection{Particle-in-cell simulation parameters}
	The simulation shown in Fig. \ref{fig:mas_sim} uses a grid resolution of \( dz = 0.002\,c/\omega_p \) in the longitudinal direction and \( dr = 0.025\,c/\omega_p \) in the radial direction, with a time step of \( dt = 1.4 \times 10^{-3}/\omega_p \). In this simulation, helium ions are mobile, and both electrons and ions are represented by 1200 particles per cell. 
	
	The simulations presented in Figs.~\ref{fig:ring_gen} and \ref{fig:mas_800nm}(b) have a grid resolution of \( dz = 0.00036\,c/\omega^\prime \) and \( dr = 0.025\,c/\omega^\prime \), with a time step of \( dt = 2.25 \times 10^{-4}/\omega{^\prime} \). In these simulations, ions are considered immobile. We point out that the resolution is given in terms of $\omega{^\prime}= ({4\pi n e^2/m_e})^{1/2}$, where $n = \SI{1.5e15}{\centi\meter^{-3}}$ is the neutral gas atomic density in Fig.~\ref{fig:ring_gen} and $n = \SI{1.6e14}{\centi\meter^{-3}}$ in Fig. \ref{fig:mas_800nm}(b). The electron plasma frequency $\omega_p$ will vary depending on how many levels are ionized. 
	
	For the simulation in Fig.~\ref{fig:ring_gen}, each ionization level is represented by 1200 particles per cell. For example, for lithium, which has three ionization levels, the maximum number of particles per cell is 3600 if all levels are fully ionized; if only the first level is ionized, there are 1200 particles per cell. In the simulation shown in Fig.~\ref{fig:mas_800nm}(b), the ionization levels are grouped into two sets: levels one to eight and levels nine to sixteen. Each group has a combined total of 1200 particles per cell, resulting in 150 particles per cell per ionization level.
	
	\subsection{Dispersion relation for the X and S waves}
	The general dielectric tensor for electrons is given by \cite{1992Stix}
	\begin{multline}
		\epsilon_{ij} = \delta_{ij} + \frac{2\pi}{\omega^2} \sum_{n=-\infty}^{\infty} \int_0^{\infty}  p_\perp^2 \, dp_\perp \int_{-\infty}^{\infty} dp_z \, I_{ij,n}(p_\perp, p_z) \\
		+ \delta_{iz}\delta_{jz} \frac{2\pi}{\omega^2} \int_0^\infty p_\perp \, dp_\perp \int_{-\infty}^\infty dp_z \, \frac{p_z}{\gamma} \left[ \frac{\partial f_0}{\partial p_z} - \frac{p_z}{p_\perp} \frac{\partial f_0}{\partial p_\perp} \right],
		\label{eq:dielectric_tensor}
	\end{multline}
	where
	\begin{multline}
		I_{ij,n} = \left[ \left( \omega - \frac{k_z p_z}{\gamma} \right) \frac{\partial f_0}{\partial p_\perp} + \frac{k_z p_\perp}{\gamma} \frac{\partial f_0}{\partial p_z} \right] \\
		\times \frac{t_{in}^*(\zeta) \, t_{jn}(\zeta)}{\gamma \omega - n \Omega_c - k_z p_z} f_0(p_\perp, p_z),
		\label{eq:I_ijn}
	\end{multline}
	and \( \zeta = k_\perp p_\perp / \Omega_c \). The functions \( t_{in} \) are defined as \( t_{xn} = n J_n(\zeta) / \zeta \), \( t_{yn} = i J_n'(\zeta) \), and \( t_{zn} = p_z J_n(\zeta) / p_\perp \), where \( J_n(\zeta) \) is the Bessel function of the first kind of order \( n \), and \( J_n'(\zeta) \) is its derivative with respect to \( \zeta \). We have assumed, without loss of generality, that the wavevector \( \mathbf{k} = k_\perp \hat{x} + k_z \hat{z} \). In Eq.~\eqref{eq:dielectric_tensor}, \( \Omega_c = e B_0 / m_e c \) is the non-relativistic cyclotron frequency, and the Lorentz factor \( \gamma \) is given by \( \gamma = \sqrt{1 + p_\perp^2 + p_z^2} \). All frequencies are normalized to the plasma frequency \( \omega_p \), wavevectors to \( \omega_p / c \), and momenta to \( m_e c \) and all dependencies on \( \gamma \) are explicit in Eq.~\eqref{eq:dielectric_tensor}. The equilibrium distribution function \( f_0(p_\perp, p_z) \) is assumed to be separable, \( f_0(p_\perp, p_z) = f_\perp(p_\perp) f_z(p_z) \), given by Eq.~\eqref{eq:f0}. 
	
	For both the X and S waves, the electric field oscillates in the plane perpendicular to the ambient magnetic field, so only the components with \( i = \{ x, y \} \) and \( j = \{ x, y \} \) are of interest. We further assume that \( \gamma \approx \gamma_\perp = \sqrt{1 + p_\perp^2} \), which is a good approximation as long as \( p_0^2 / 2 \gg T_z \), where \( p_0 \) is the ring momentum radius and \( T_z \) is the longitudinal temperature (normalized to \( m_e c^2 \)).
	
	\textit{X waves}---For the X waves, we set \( k_z = 0 \) in Eq.~\eqref{eq:dielectric_tensor}. The integration over \( p_z \) becomes trivial since \( f_z(p_z) \) integrates to unity (\( \int_{-\infty}^\infty f_z(p_z) \, dp_z = 1 \)). Therefore, the dielectric tensor simplifies to
	\begin{equation}
		\epsilon_{ij} = \delta_{ij} + \frac{2\pi}{\omega^2} \sum_{n=-\infty}^{\infty} \int_0^{\infty} \frac{p_\perp^2 \, t_{in}^*(\zeta) t_{jn}(\zeta)}{\gamma - n \Omega_c / \omega} \frac{\partial f_\perp}{\partial p_\perp} \, dp_\perp, \label{eq:die_X}
	\end{equation}
	where \( \zeta = k_\perp p_\perp / \Omega_c \). We note that in this case, the growth rate of the X waves is independent of \( T_z \) due to the assumption \( p_0^2 / 2 \gg T_z \). Additionally, the unstable spectrum is only significantly affected by the value of $T_\perp$ when it has values $T_\perp \gtrsim \SI{1}{\kilo eV}$. In the laser-ionized plasmas, this value is typically a few hundred \SI{}{eV}, thus one can use $f_\perp(p_\perp) = (2\pi p_\perp)^{-1} \delta(p_\perp -p_0)$. 
	
	Using Eq.~\eqref{eq:die_X}, the dispersion relation can be calculated from the dispersion tensor \( D_{ij} = k_i k_j - k^2 \delta_{ij} + \omega^2 \epsilon_{ij} \) as
	\begin{equation}
		D_{xx} D_{yy} - D_{xy} D_{yx} = 0, \label{eq:disp2}
	\end{equation}
	where
	\begin{multline}
		D_{ij} = (\omega^2 - k^2) \delta_{ij} + k_i k_j \\+ \sum_{n=-\infty}^{\infty}\left[\frac{\Lambda_{ij,n}\,\omega^2}{(\omega - n\Omega_c / \gamma_0)^2} + \frac{\Xi_{ij,n}\,\omega}{\omega - n\Omega_c / \gamma_0}\right]. \label{eq:dij}
	\end{multline}
	Here, \( \gamma_0 = \sqrt{1 + p_0^2} \) and the coefficients are given by
	\begin{align*}
		\Lambda_{xx,n} &= \frac{n^2 p_0^2}{\gamma_0^3 \zeta^2} J_n^2(\zeta), \quad \Lambda_{yy,n} = \frac{p_0^2}{\gamma_0^3} \left[ J_n'(\zeta) \right]^2, \\
		\Lambda_{xy,n} &= i \frac{n p_0^2}{\gamma_0^3 \zeta} J_n(\zeta) J_n'(\zeta), \quad
		\Xi_{xx,n} = -\frac{2 n^2}{\gamma_0 \zeta} J_n(\zeta) J_n'(\zeta), \\ \Xi_{yy,n} &= -\frac{2}{\gamma_0} J_n'(\zeta) \left[ \zeta J_n''(\zeta) + J_n'(\zeta) \right], \\
		\Xi_{xy,n} &= -i \frac{n}{\gamma_0} \left[ \frac{1}{\zeta} J_n(\zeta) J_n'(\zeta) + \left[ J_n'(\zeta) \right]^2 + J_n(\zeta) J_n''(\zeta) \right],
	\end{align*}
	with \( \Lambda_{yx,n} = \Lambda_{xy,n}^* \) and \( \Xi_{yx,n} = \Xi_{xy,n}^* \), where the superscript \( * \) denotes complex conjugation.
	
	Although the sum in Eq.~\eqref{eq:dij} formally extends over all cyclotron harmonics \( n \), in practice, the solutions of Eq.~\eqref{eq:disp2} have frequencies close to the harmonics (\( \omega \approx n \Omega_c / \gamma_0 \)). Therefore, only the resonant terms (those with \( n \) corresponding to the harmonic of interest) significantly contribute to the sum. Nevertheless, our results in Fig. \ref{fig:mas_theory} include all harmonics between \( n \in [-3,\, 3]\).
	
	\textit{S waves}---For the S waves, we set \( k_\perp = 0 \) in Eq.~\eqref{eq:dielectric_tensor}. After integrating over \( p_z \), the dielectric tensor becomes
	\begin{multline}
		\epsilon_{ij} = \delta_{ij} + \frac{2\pi}{\omega^2}\sum_{n=-\infty}^{\infty} t_{in}^*(0) t_{jn}(0) \times\\ \left\{ \int_0^{\infty} \left(\frac{1}{T_z}\frac{ p_\perp^3}{\gamma_\perp } - \frac{n\Omega_c\omega}{T_zk_z^2}\frac{p_\perp^3}{\gamma_\perp^2} \right)\left[1+\xi_nZ(\xi_n)\right] f_\perp dp_\perp \right.\\\left.+ \int_0^{\infty} \left(\frac{p_\perp^3}{\gamma_\perp^3} -2\frac{p_\perp}{\gamma_\perp}\right)\left[1-\frac{n\Omega_c}{\sqrt{2T_z}k_z}Z(\xi_n)\right]f_\perp dp_\perp \right\},
	\end{multline}
	where \( \xi_n = (\gamma_\perp \omega - n \Omega_c) / (\sqrt{2 T_z} k_z) \) and \( Z(\xi) \) is the plasma dispersion function \cite{1961Fried} defined by
	\begin{equation}
		Z(\xi) = \frac{1}{\sqrt{\pi}} \int_{-\infty}^{\infty} \frac{e^{-u^2}}{u - \xi} \, du.
	\end{equation}
	
	For \( k_\perp = 0 \), only \( n = \pm 1 \) contribute to the sum. The dielectric tensor simplifies to
	\begin{equation}
		\epsilon_{ij} = \delta_{ij} + \frac{1}{\omega^2} \left[ t_{i1}^*(0) t_{j1}(0) \mathcal{I}_1 + t_{i-1}^*(0) t_{j-1}(0) \mathcal{I}_{-1} \right],
	\end{equation}
	where
	\begin{multline}
		\mathcal{I}_n =\int_0^{\infty} \left(\frac{1}{T_z}\frac{ p_\perp^3}{\gamma_\perp } - \frac{n\Omega\omega}{T_zk_z^2}\frac{p_\perp^3}{\gamma_\perp^2} \right)\left[1+\xi_nZ(\xi_n)\right] f_\perp dp_\perp\\+ \int_0^{\infty} \left(\frac{p_\perp^3}{\gamma_\perp^3} -2\frac{p_\perp}{\gamma_\perp}\right)\left[1-\frac{n\Omega}{\sqrt{2T_z}k_z}Z(\xi_n)\right]f_\perp dp_\perp. \label{eq:I_n}
	\end{multline}
	From the dispersion relation \( D_{xx} D_{yy} - D_{xy} D_{yx} = 0 \), we obtain two possible solutions:
	\begin{align}
		&\omega^2 - k_z^2 + \frac{1}{2} \mathcal{I}_1 = 0, \label{eq:Rwave_dispersion}\\
		&\omega^2 - k_z^2 + \frac{1}{2} \mathcal{I}_{-1} = 0,
	\end{align}
	representing right-handed (R) and left-handed (L) circularly polarized waves, respectively. We focus on the R waves described by Eq.~\eqref{eq:Rwave_dispersion}.
	
	By performing the integration in Eq.~\eqref{eq:I_n} for \( n = 1 \), Eq.~\eqref{eq:Rwave_dispersion} becomes
	\begin{multline}
		\omega^2 - k_z^2 + \frac{p_0^2}{2 \gamma_0 T_z} \left( 1 - \frac{\Omega_c \omega}{\gamma_0 k_z^2} \right) \left[ 1 + \xi Z(\xi) \right] \\
		+ \left( \frac{p_0^2}{2 \gamma_0^3} - \frac{1}{\gamma_0} \right) \left[ 1 - \frac{\Omega_c}{\sqrt{2 T_z} k_z} Z(\xi) \right] = 0, \label{eq:disp_Tz_kz}
	\end{multline}
	where \( \xi = (\gamma_0 \omega - \Omega_c)/(\sqrt{2 T_z} k_z) \).
	
	\textit{Numerical solutions}---Equation~\eqref{eq:disp2} for the X waves can be transformed into a polynomial equation; however, the coefficients can be large and lead to numerical instability. Equation~\eqref{eq:disp_Tz_kz} for the S waves is a transcendental equation due to the plasma dispersion function \( Z(\xi) \). To solve both dispersion relations accurately, we employ numerical methods; specifically Newton's method.

\providecommand{\noopsort}[1]{}\providecommand{\singleletter}[1]{#1}%

\end{document}